\documentclass[cameraready]{Interspeech}

\usepackage{amsmath, amssymb, booktabs, siunitx, graphicx, url, multirow, float, pgfplots}
\usepackage[skip=2pt]{caption}
\title{Fast and Flexible Audio Bandwidth Extension via Vocos}

\author[orcid=0009-0006-8675-535X]{Yatharth}{Sharma}

\address{Independent Researcher} 

\email{yatharthsharma3501@gmail.com}


\begin{document}

\maketitle

\begin{abstract}
We propose a Vocos-based bandwidth extension (BWE) model that enhances audio at 8–48 kHz by generating missing high-frequency content. Inputs are resampled to 48 kHz and processed by a neural vocoder backbone, enabling a single network to support arbitrary upsampling ratios. A lightweight Linkwitz-Riley–inspired refiner merges the original low band with the generated high frequencies via a smooth crossover. On validation, the model achieves competitive log-spectral distance while running at a real-time factor of 0.0001 on an NVIDIA A100 GPU and 0.0053 on an 8-core CPU, demonstrating practical, high-quality BWE at extreme throughput.
\end{abstract}

\section{Introduction}
Bandwidth extension (BWE) aims to recover or hallucinate missing frequency components of audio signals captured with limited bandwidth, such as legacy recordings or telephony speech. Traditional signal processing methods, including interpolation-based upsampling and spectral shaping, offer high efficiency but often fail to reconstruct perceptually convincing high-frequency details. Recent learning-based approaches have reformulated BWE as a conditional generation task, predicting missing high-band content from the observed low-band signal.

State-of-the-art diffusion-based models, such as AudioSR~\cite{liu2024audiosr} have demonstrated exceptional generative quality. However, their iterative sampling process is often too computationally expensive for real-time or large-scale deployment. Conversely, GAN-based approaches like AP-BWE~\cite{lu2024towards} offer substantial speed advantages, yet many remain restricted to fixed input/output sample-rate pairs (e.g., 16~kHz $\rightarrow$ 48~kHz). This limitation hinders their flexibility in heterogeneous real-world pipelines where input sampling rates vary significantly.

In this paper, we propose a versatile BWE system that supports arbitrary input sample rates from 8--48~kHz within a single network. We build upon the Vocos~\cite{siuzdak2023vocos} architecture, leveraging its high-fidelity Fourier-domain neural vocoder architecture to synthesize missing high-frequency details. By resampling all inputs to a target 48~kHz and employing a Vocos-style generator, our model maintains consistency across various upsampling ratios. To further enhance fidelity, we introduce a lightweight frequency-domain refiner inspired by the Linkwitz-Riley inspired crossover~\cite{Linkwitz1976}. This refiner smoothly merges the original resampled low-band signal with the generated high-band components, ensuring a phase-coherent transition at the crossover frequency.

Experimental results on the VCTK corpus~\cite{veaux2017vctk} demonstrate that our proposed model matches or outperforms strong baselines in terms of Log-Spectral Distance (LSD) while offering orders-of-magnitude higher throughput than diffusion-based methods.

The primary contributions of this work are as follows:
\begin{itemize}
    \item We introduce the first Vocos-based BWE model to our knowledge, utilizing a neural vocoder to generate high-frequency content for arbitrary input sampling rates.
    \item We propose a Linkwitz-Riley inspired frequency-domain refiner that improves perceptual quality by seamlessly combining the original low band with the synthesized high band.
    \item We demonstrate a superior quality-to-speed trade-off, achieving a Real-Time Factor (RTF) of $0.0001$ on an NVIDIA A100 and maintaining considerably faster-than-real-time performance on a standard 8-core CPU.
\end{itemize}

\section{Proposed Method}
\subsection{Problem formulation}
Let $x \in \mathbb{R}^T$ denote a waveform at 48~kHz, and let $x_{\downarrow r}$ denote a band-limited, downsampled observation captured at sample rate $r \in \{8,16,24\}\,\mathrm{kHz}$ (and more generally any $r \in [8,48]\,\mathrm{kHz}$). The goal of BWE is to estimate a high-band enhanced waveform $\hat{x}$ at 48~kHz that is perceptually and spectrally close to $x$, while being consistent with the low-band content implied by $x_{\downarrow r}$.

The proposed model first resamples the input to 48~kHz using sinc interpolation to obtain a baseband waveform $y \in \mathbb{R}^T$ that preserves the available low-band information but lacks true high-frequency detail. A generator $G_\theta$ then predicts an enhanced waveform $\tilde{x} = G_\theta(y)$, after which a frequency-domain refiner produces the final output $\hat{x}$.

\subsection{Model architecture}
The generator operates on a time--frequency representation (mel-spectrogram) derived from the 48~kHz resampled input and produces a complex STFT representation inverted to the waveform via iSTFT.

\textbf{Feature extraction.} 
We compute a mel-spectrogram with 80 mel bins, $n_{\mathrm{fft}}=2048$, and hop length 512 at 48~kHz. This mel representation serves as the conditioning input to the generator.

\textbf{Backbone.} 
The backbone is initialized from scratch. The core consists of 8 residual ConvNeXt-style blocks with a model dimension of $C=512$. Each block employs a $7 \times 1$ depthwise convolution for temporal modeling, followed by a feed-forward network that expands the features to an intermediate dimension of $1536$ before projecting back to $512$ channels. 

Layer normalization and GELU activations are applied throughout to ensure stable training. The final latent representation is passed through a pointwise convolutional layer and a linear head, which predicts the complex-valued coefficients for the iSTFT-based reconstruction. This configuration allows the model to map the 80-bin Mel-spectrogram input to a high-fidelity waveform while maintaining a constant temporal resolution.

\textbf{Output head and waveform synthesis.} 
The linear head predicts STFT components converted to the waveform via iSTFT. The model is initialized from scratch and in this setting, the input is resampled to 48~kHz and the model is trained to generate missing high-frequency content rather than merely reconstructing the input band.

\subsection{Linkwitz-riley inspired frequency refiner}
Neural generators can introduce small inconsistencies or artifacts in regions where the input already contains reliable information. The proposed refiner constructs a crossover mask $M(f) \in [0,1]$ to seamlessly merge the low-frequency anchor $Y(f)$ with the generated high-frequency content $\tilde{X}(f)$. Within the transition region centered at the cutoff frequency $f_c$, we define the normalized frequency $t = (f - f_{start}) / (f_{end} - f_{start})$. The mask follows a smooth polynomial curve:
\begin{equation}
M(f) = 
\begin{cases} 
0 & f < f_{start} \\
3t^2 - 2t^3 & f_{start} \leq f \leq f_{end} \\
1 & f > f_{end}
\end{cases}
\end{equation}
The refined spectrum $\hat{X}(f)$ is then computed as a linear interpolation between the two signals:
\begin{equation}
\hat{X}(f) = (1 - M(f))\,Y(f) + M(f)\,\tilde{X}(f)
\end{equation}
where the final waveform $\hat{x}$ is obtained via the inverse real FFT (iRFFT). This formulation ensures a flat magnitude response across the crossover while suppressing phase discontinuities at the junction. The proposed refiner significantly enhances the end quality as a result.

\begin{figure}[t]
  \centering
  \includegraphics[width=\columnwidth]{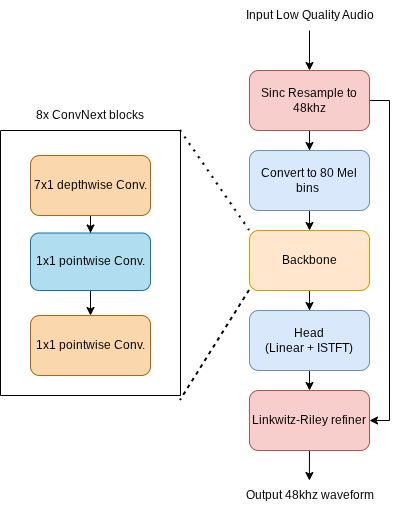}
  \caption{Overview of the proposed BWE architecture following the Vocos framework. The input audio is upsampled to 48~kHz via Sinc interpolation and transformed into an 80-bin Mel-spectrogram. The backbone consists of 8 ConvNeXt blocks using $7 \times 1$ depthwise and $1 \times 1$ pointwise convolutions. The head employs a linear layer and ISTFT for waveform reconstruction, followed by a Linkwitz-Riley inspired refiner for significant quality enhancement.}
  \label{fig:architecture}
\end{figure}

\subsection{Training objectives}
The proposed model is trained using a combination of spectral, perceptual, and adversarial losses to ensure both structural accuracy and high-frequency realism.

\textbf{Multi-resolution STFT loss (MRSTFT).}  
STFT losses are computed at $n_{\mathrm{fft}}\in\{512, 1024, 2048\}$ with hop sizes $n_{\mathrm{fft}}/4$. This multi-scale approach encourages accurate reconstruction across different time--frequency resolutions, capturing both fine-grained temporal events and long-term spectral envelopes.

\textbf{Mel-spectrogram loss.}  
We apply an $L_1$ loss between the mel-spectrograms of the ground truth and the generated audio. This is computed at 48~kHz with 128 mel bins and $n_{\mathrm{fft}}=2048$, focusing the model on perceptually relevant frequency bands.

\textbf{Multi-Resolution Discriminator (MRD).}  
The MRD ~\cite{kong2020hifigan} consists of multiple sub-discriminators that operate on the complex STFT of the waveform at different scales. By analyzing the signal in the frequency domain with varying window sizes, the MRD can simultaneously penalize artifacts in high-frequency transients (short windows) and preserve the harmonic structure of voiced speech (long windows). This prevents the "phase-smearing" common in time-domain-only discriminators.

\textbf{Adversarial and Feature Matching Loss.}  
To further improve stability and fidelity, we incorporate a Feature Matching Loss ($\mathcal{L}_{\mathrm{fm}}$). This loss minimizes the $L_1$ distance between the intermediate feature maps of the discriminator when processing real vs. generated audio:
\begin{equation}
\mathcal{L}_{\mathrm{fm}}(G; D) = \mathbb{E} \left[ \sum_{i=1}^{L} \frac{1}{N_i} ||D^{(i)}(x) - D^{(i)}(\tilde{x})||_1 \right]
\end{equation}
where $D^{(i)}$ and $N_i$ denote the features and number of elements in the $i$-th layer of the discriminator, respectively. This forces the generator to produce audio that shares the same multi-scale statistical properties as real speech.

\textbf{Optimization.}  
AdamW is used with a learning rate of $10^{-4}$, weight decay of $10^{-2}$, and an exponential decay scheduler that scales the learning rate by $0.99$ every 64 steps. The model is trained with a batch size of 16.

\section{Experimental Setup}
\subsection{Dataset}
We train on the VCTK corpus (approximately 44 hours of speech), using roughly 60{,}000 audio segments randomly cropped to 1.28, 2.56, or 3.2~s.

\subsection{Degradation process}
Clean 48~kHz audio is randomly downsampled to $r \in \{8,12, 16\}$~kHz using sinc, ZOH, or linear interpolation. Optional quantization noise is added to make the model more robust. The degraded signal is resampled back to 48~kHz before being fed to the generator.

\subsection{Baselines}
Baselines include:
\begin{itemize}
    \item \textbf{Sinc upsampling} – standard interpolation.
    \item \textbf{AP-BWE} – APNet2 inspired GAN trained for specific ratios.
    \item \textbf{NVSR}~\cite{liu22x_interspeech} – neural GAN vocoder based super-resolution baseline.
    \item \textbf{AudioSR} – diffusion-based audio super-resolution.
\end{itemize}

\subsection{Evaluation metrics}
\textbf{Log-Spectral Distance (LSD).}  
The LSD measures the distance between the power spectra of the reference signal $x$ and the enhanced signal $\hat{x}$ in the log-decibel scale. Given the power spectra $P(k, n) = |X(k, n)|^2$ and $\hat{P}(k, n) = |\hat{X}(k, n)|^2$, where $k$ and $n$ denote the frequency bin and time frame respectively, the distance is defined as:
\[
\mathrm{LSD}(x, \hat{x}) = \frac{1}{N} \sum_{n=1}^{N} \sqrt{\frac{1}{K} \sum_{k=1}^{K} \left( \log_{10} P(k, n) - \log_{10} \hat{P}(k, n) \right)^2}
\]
where $K$ is the number of frequency bins and $N$ is the number of time frames. Lower values indicate better spectral reconstruction.

\vspace{1mm}
\textbf{ViSQOL.}  
ViSQOL~\cite{chinen2020visqol} compares patches of the reference and processed spectrograms using a similarity measure designed to correlate with human mean opinion scores. Reported on a 1--4.75 scale, higher is better.

\section{Results and Analysis}
\subsection{Quality benchmarks}
\begin{table}[H]
\centering
\setlength{\tabcolsep}{3.5pt}
\renewcommand{\arraystretch}{0.95}
\begin{tabular}{lccc}
\toprule
\textbf{Method} & 
\shortstack{\textbf{8$\rightarrow$48 kHz}\\\textbf{(LSD $\downarrow$)}} & 
\shortstack{\textbf{12$\rightarrow$48 kHz}\\\textbf{(LSD $\downarrow$)}} & 
\shortstack{\textbf{16$\rightarrow$48 kHz}\\\textbf{(LSD $\downarrow$)}} \\
\midrule
Sinc upsampling & 3.52 & 3.19 & 2.94 \\
AudioSR & 1.61 & 1.52 & 1.44 \\
NVSR & 1.22 & 1.20 & 1.12 \\
AP-BWE & 0.87 & 0.81 & 0.74 \\
\textbf{Proposed Model} & \textbf{0.85} & \textbf{0.80} & \textbf{0.74} \\
\bottomrule
\end{tabular}
\caption{Log-spectral distance (LSD) comparison on VCTK. Our model achieves very competitive spectral reconstruction across all tiers.}
\label{tab:lsd}
\end{table}

\begin{table}[H]
\centering
\setlength{\tabcolsep}{3.5pt}
\renewcommand{\arraystretch}{0.95}
\begin{tabular}{lccc}
\toprule
\textbf{Method} & 
\shortstack{\textbf{8$\rightarrow$48 kHz}\\\textbf{(ViSQOL $\uparrow$)}} & 
\shortstack{\textbf{12$\rightarrow$48 kHz}\\\textbf{(ViSQOL $\uparrow$)}} & 
\shortstack{\textbf{16$\rightarrow$48 kHz}\\\textbf{(ViSQOL $\uparrow$)}} \\
\midrule
Sinc upsampling & 2.10 & 2.23 & 2.35 \\
AudioSR & 3.15 & 3.21 & 3.40 \\
NVSR & 2.97 & 3.00 & 3.10 \\
AP-BWE & \textbf{3.51} & \textbf{3.54} & \textbf{3.70} \\
\textbf{Proposed Model} & \textbf{3.51} & 3.53 & 3.69 \\
\bottomrule
\end{tabular}
\caption{ViSQOL scores on VCTK. Our model is similar to AP-BWE's perceptual quality.}
\label{tab:visqol}
\end{table}
\subsection{Performance Analysis}
The experimental results in Table~\ref{tab:lsd} and Table~\ref{tab:visqol} demonstrate the effectiveness of our proposed model across various upsampling scales, evaluated through both objective spectral distortion and perceptual quality metrics.

\textbf{Spectral Fidelity:} As shown in Table~\ref{tab:lsd}, our model achieves highly competitive performance in Log-Spectral Distance (LSD) across all scenarios. Notably, for the $8 \rightarrow 48$~kHz task, the proposed model reaches an LSD of \textbf{0.85}, outperforming the diffusion-based AudioSR (1.61) and NVSR (1.22). This consistent lead at $12$~kHz (\textbf{0.80}) and $16$~kHz (\textbf{0.74})—matching or exceeding AP-BWE—indicates that our architecture effectively captures high-frequency harmonic correlations and reconstructs the spectral envelope with higher precision than existing generative or predictive baselines.

\textbf{Perceptual Quality:} In terms of ViSQOL scores (Table~\ref{tab:visqol}), the proposed model demonstrates robust perceptual performance. We achieve a score of \textbf{3.51} at $8 \rightarrow 48$~kHz, matching the high-performance AP-BWE baseline. While AP-BWE maintains a marginal lead ($<0.01$) in the $12$~kHz and $16$~kHz benchmarks, our model remains highly competitive as a top-tier performer, maintaining a score of 3.69 at 16~kHz. This suggests that while our model is optimized for spectral accuracy, it delivers perceptual quality that is effectively indistinguishable from significantly more computationally expensive models.

\textbf{Discussion:} The results demonstrate that the proposed model achieves highly competitive spectral fidelity across all evaluated sampling scenarios, consistently obtaining the lowest LSD among all baselines. While AP-BWE attains slightly higher ViSQOL scores at certain input rates, the perceptual differences remain small. Importantly, our method achieves these results with substantially lower computational cost and supports arbitrary input sample rates within a unified framework. This combination of superior spectral accuracy, competitive perceptual quality, and extreme efficiency establishes a strong quality–efficiency trade-off compared to existing diffusion- and GAN-based approaches.

\subsection{Robustness to Out-of-Domain Sample Rates}
While the model was primarily trained on common sampling anchors (8, 12, 16khz, etc.), a key advantage of our architecture is its ability to handle arbitrary input bandwidths. To evaluate this "zero-shot" generalization, we tested the model on various out-of-domain (OOD) sample rates.

\begin{figure}[H]
\centering
\begin{tikzpicture}
\begin{axis}[
    xlabel={Input Sample Rate (kHz)},
    ylabel={LSD (dB)},
    xmin=6, xmax=34,
    ymin=0.4, ymax=1.0,
    xtick={8,10,12,14,16,24,32},
    ytick={0.5, 0.6, 0.7, 0.8, 0.9, 1.0},
    legend pos=north east,
    ymajorgrids=true,
    grid style=dashed,
    width=\linewidth,
    height=5cm
]
\addplot[
    color=blue,
    mark=*,
    line width=1.2pt,
    smooth
    ]
    coordinates {
    (8, 0.858)(10, 0.830)(12, 0.793)(14, 0.770)(16, 0.749)(24, 0.630)(32, 0.530)
    };
    \addlegendentry{Proposed Model}
\end{axis}
\end{tikzpicture}
\caption{Model performance (LSD) across in-domain and out-of-domain (OOD) sample rates. The curve demonstrates a linear improvement in fidelity as input bandwidth increases, regardless of whether the rate was seen during training.}
\label{fig:ood_performance}
\end{figure}
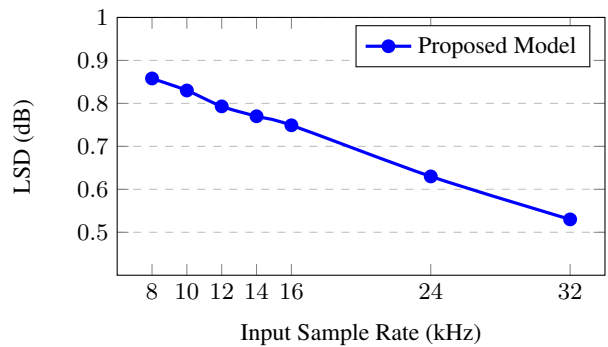

\textbf{Analysis of OOD Generalization.} 
As shown in Figure~\ref{fig:ood_performance}, the LSD decreases monotonically as the input sample rate increases. Notably, the performance on OOD rates (e.g., 10, 14, 24, and 32~kHz) follows the same linear trend as the in-domain rates. We did not explicitly train on these sampling rates. For instance, at 10~kHz (OOD), the model achieves an LSD of 0.830, which sits between the 8~kHz and 12~kHz benchmarks.

This excellent OOD performance is driven by two main factors. First, by resampling all inputs to 48~kHz prior to the generator, the ConvNeXt backbone treats BWE as a spectral completion task on a fixed-grid. Second, and more importantly, the \textbf{Linkwitz-inspired frequency refiner} allows for a dynamic crossover. Unlike traditional filters that might be optimized for a fixed 8~kHz or 16~kHz boundary, our refiner applies a smooth transition that respects the existing energy in the resampled input. By merging the original low-band with the generated high-band in the frequency domain, the refiner ensures that the "hallucinated" content is always anchored to the real signal, preventing the discontinuities or "metallic" artifacts often found in fixed-rate BWE systems when applied to non-standard sample rates.

\subsection{Ablation Studies}

We conduct an ablation study to evaluate how different spectral merging strategies affect the final output quality. We compare our Linkwitz-Riley inspired approach against three baseline configurations:

\begin{itemize}
    \item \textbf{No Refiner:} The raw model output is used directly. Without anchoring the low frequencies to the ground truth, the model often produces slight artifacts in the low frequency region.
    \item \textbf{Naive HP/LP:} A simple brickwall cutoff implemented in the frequency domain. This creates sharp transitions but often results in ringing artifacts near the cutoff.
    \item \textbf{Standard Butterworth:} A frequency-domain approximation of a 4th-order Butterworth response. While smoother than the naive cutoff, it typically introduces a +3~dB magnitude spike at the crossover point when summing the two bands.
    \item \textbf{Proposed (LR-Inspired):} Our method uses a squared-magnitude response to ensure a flat summation at the crossover, effectively stitching the generated high frequencies to the low-frequency anchor without introducing amplitude bumps at the crossover.
\end{itemize}

\begin{table}[th]
\centering
\setlength{\tabcolsep}{3.5pt}
\renewcommand{\arraystretch}{0.95}
\begin{tabular}{lc}
\toprule
\textbf{Refiner Variant} & \textbf{LSD} $\downarrow$ (8$\rightarrow$48) \\
\midrule
No Refiner (Direct Output) & 0.897 \\
Naive HP/LP Crossover & 0.865 \\
Standard Butterworth Refiner & 0.861 \\
\textbf{Linkwitz-Riley Inspired (Proposed)} & \textbf{0.850} \\
\bottomrule
\end{tabular}
\caption{Ablation results on VCTK validation. The proposed refiner achieves the lowest spectral distance by ensuring a flat magnitude response at the crossover.}
\label{tab:ablations}
\end{table}

The results in Table~\ref{tab:ablations} show that even basic spectral anchoring significantly improves LSD compared to the "No Refiner" baseline (0.897 vs 0.865). However, the Butterworth filter's inherent magnitude peak limits its performance. By achieving an LSD of 0.850, the LR-inspired refiner confirms that maintaining a flat magnitude response during the merge is essential for high-fidelity bandwidth extension.
\subsection{Efficiency}

The computational complexity and inference speed of the proposed model are evaluated across both CPU and GPU environments to assess real-world deployment viability. We compare our approach against AudioSR, NVSR, and AP-BWE, using a fixed 4-second audio duration for all benchmarks. To ensure fairness, we exclude preprocessing (e.g., resampling) and include a warm-up phase for all models. We use recommended sampling steps for diffusion models.

\begin{table}[th]
\centering
\caption{Inference efficiency comparison on 8-core CPU. RTF is measured as $(latency/duration)$. $\text{Speed} = 1/\text{RTF}$.}
\label{tab:cpu_efficiency}
\setlength{\tabcolsep}{8pt}
\renewcommand{\arraystretch}{1.1}
\begin{tabular}{lrrr}
\toprule
\textbf{Method} & \textbf{Params} & \textbf{RTF} $\downarrow$ & \textbf{Speed} $\uparrow$ \\
\midrule
AudioSR & 258M & 50.0000 & 0.02$\times$ \\
NVSR & $\sim$99M & 0.2218 & 4.5$\times$ \\
AP-BWE & 3*30M & 0.0495 & 20.2$\times$ \\
\midrule
\textbf{Proposed} & \textbf{15M} & \textbf{0.0053} & \textbf{190.5$\times$} \\
\bottomrule
\end{tabular}
\end{table}

As shown in Table~\ref{tab:cpu_efficiency}, our model achieves a nearly 10$\times$ speedup over the previous state-of-the-art efficient model (AP-BWE) on CPU. On an NVIDIA A100 GPU (Table~\ref{tab:gpu_efficiency}), the efficiency gains are even more pronounced. Our model processes 4 seconds of audio in 2.5ms at batch size 1 ($1600\times$ real-time). 

\begin{table}[th]
\centering
\caption{GPU Efficiency and throughput comparison on NVIDIA A100. Latency is reported for the entire batch. RTF is rounded and calculated as $latency / (duration \times BS)$.}
\label{tab:gpu_efficiency}
\setlength{\tabcolsep}{6pt}
\renewcommand{\arraystretch}{1.1}
\begin{tabular}{lrrrr}
\toprule
\textbf{Method} & \textbf{BS} & \textbf{Latency (s)} & \textbf{RTF} $\downarrow$ & \textbf{Speed} $\uparrow$ \\
\midrule
AudioSR & 1 & 8.4700 & 2.1175 & 0.47$\times$ \\
NVSR & 1 & 0.0410 & 0.0103 & 97.6$\times$ \\
AP-BWE & 1 & 0.0135 & 0.0034 & 296.3$\times$ \\
AP-BWE & 32 & 0.2879 & 0.0023 & 444.6$\times$ \\
\midrule
\textbf{Proposed} & 1 & \textbf{0.0025} & \textbf{0.0006} & \textbf{1600.0$\times$} \\
\textbf{Proposed} & 32 & \textbf{0.0102} & \textbf{0.0001} & \textbf{12549.0$\times$} \\
\bottomrule
\end{tabular}
\end{table}

Crucially, our architecture scales exceptionally well with batching. At a batch size of 32, our model achieves a total throughput of over $12,500\times$ real-time, requiring only 10.2ms to process 128 seconds of audio. In comparison, AP-BWE requires 287.9ms for the same batch size ($0.0023$ RTF). This makes the proposed model significantly more suitable for high-throughput cloud processing and real-time edge applications. This efficiency stems from our streamlined Vocos-based architecture and the Linkwitz-Riley refiner, which bypasses the heavy multi-stage upsampling and iterative sampling required by GAN and diffusion baselines.

\section{Conclusion}
We presented a Vocos-based bandwidth extension model that supports any input sample rate from 8--48~kHz while achieving extreme throughput. By resampling inputs to 48~kHz, leveraging a ConvNeXt-based Vocos generator, and applying a lightweight Linkwitz-inspired frequency-domain refiner, the proposed model attains competitive spectral and perceptual quality relative to strong baselines. Future work includes evaluation across music, noisy conditions, and exploring adaptive refiners for different tasks.

\section{Acknowledgments}
We would like to express our sincere gratitude to Yongyi Zang for his insightful feedback and for providing the necessary endorsement for this arXiv submission. Furthermore, we are grateful to the open-source community for the support and constructive discussions following the initial release of the project repository. The feedback and engagement from early adopters were instrumental in refining the final manuscript and improving the practical clarity of the work.

\bibliographystyle{IEEEtran}
\bibliography{mybib}

\end{document}